\documentclass[preprintnumbers,prd,twocolumn,showpacs,floatfix,preprintnumbers,letterpaper,amsmath,amssymb,nofootinbib,superscriptaddress]{revtex4}
\usepackage{graphicx}
\usepackage{epsfig}
\usepackage{bm}
\usepackage{amsfonts}
\usepackage{color}
\usepackage{subfigure}

\usepackage{amsmath,amssymb}

\begin{document}

\title{ Inflationary generalized Chaplygin gas and general dark
energy in the light of the Planck and BICEP2 experiments}

\author{Bikash R Dinda}
\email{bikash@ctp-jamia.res.in}
\affiliation{Centre for Theoretical Physics, Jamia Millia Islamia,
New Delhi-110025, India}

\author{Sumit Kumar}
\email{sumit@ctp-jamia.res.in}
\affiliation{Centre for Theoretical Physics, Jamia Millia Islamia,
New Delhi-110025, India}

\author{Anjan A Sen}
\email{aasen@jmi.ac.in}
\affiliation{Centre for Theoretical Physics, Jamia Millia Islamia,
New Delhi-110025, India}

\begin{abstract}
In this work, we study an inflationary scenario in the presence of Generalized Chaplygin Gas (GCG). We show that in Einstein gravity, GCG is not a suitable candidate for inflation; but in a five dimensional brane world scenario, it can work as a viable inflationary model. We calculate the relevant quantities such as $n_{s}$, $r$ and $A_{s}$ related to the primordial scalar and tensor fluctuations, and using their recent bounds from Planck and BICEP2, we constrain the model parameters as well as the five-dimensional Planck mass. But as a slow-roll inflationary model with a power-law type scalar primordial power spectrum, GCG as an inflationary model can not resolve the tension between results from BICEP2 and Planck with a concordance $\Lambda$CDM Universe. We show that going beyond the concordance $\Lambda$CDM model and  incorporating more general dark energy behaviour, this tension may be eased. We also obtain the constraints on the $n_{s}$ and $r$ and the GCG model parameters using Planck+WP+BICEP2 data considering the CPL dark energy behaviour.
\end{abstract}

\maketitle

\date{\today}

\maketitle

\section{Introduction}
The present cosmological observations are amazingly consistent with a Universe which has two accelerated expansion phases in its entire evolution history from big bang till today. One of these accelerating epochs presumably occurred during the early phase of the cosmological evolution when the energy scale of the Universe was close to the Planck scale. This accelerating period was first proposed around 1980 in order to solve the puzzles like flatness, horizon and monopole problems in standard cosmology \cite{inf}. An epoch of exponential expansion was proposed in order to solve these problems. Although a simple cosmological constant can give rise to such exponential expansion, the Universe can never exit from this accelerating phase in such a scenario and hence does not enter into a decelerated phase which is necessary for subsequent processes like nucleosynthesis, structure formation. To solve this exit problem, a scalar field theory was introduced where the field rolls over a sufficiently flat potential (slow 
rolling) and can mimic a cosmological constant like behaviour. Such a scalar field (inflaton) can drive a near exponential expansion. The exit from the inflationary era is ensured as the scalar field reaches the non-flat region (fast roll phase) of the potential.
One of the greatest successes of inflationary model is the generation of primordial density fluctuations in the Universe which can act as the seeds for the large scale inhomogeneities that is necessary for the structure formation of the Universe \cite{pert}. The quantum fluctuations of  the inflaton during inflation can produce such primordial density fluctuations. Given any scalar field inflationary model, one can calculate the spectrum of this primordial density fluctuation which is essentially related to the scalar part of the metric fluctuations. Moreover, one can also calculate the tensor fluctuations in the metric produced during inflation which result a stochastic gravitational wave background on large cosmological scales. Cosmologically both these primordial fluctuations are interesting as they produce observable features in the temperature anisotropy of the cosmic microwave background radiation (CMBR).  Hence measuring temperature anisotropy in CMBR enables us to constrain the spectrum of these 
initial fluctuations which in turn can constrain the gravitational physics close to Planck scale.  The scalar part of the temperature anisotropy was first measured by COBE \cite{cobe} in early nineties and subsequently by a host of CMBR experiments e.g BOOMERANG \cite{boom}, WMAP \cite{wmap} and more recently by Planck \cite{planck}. For the Polarization in CMBR, DASI \cite{dasi} first detected the E-mode polarization in CMBR in 2001. But the B-mode polarization in CMBR which is a clear evidence for the existence of primordial gravitational waves generated through the tensor fluctuations during inflation had not been detected until recently. But just recently, the BICEP2 experiment \cite{bicep2} has announced the detection of the B-mode polarization signal in CMBR ruling out the zero tensor fluctuation at $7\sigma$ confidence level. This is an extraordinary result for cosmology and if confirmed by future polarization data from Planck satellite, it will establish the fact that there was an accelerating epoch 
in the Universe prior to the radiation era. From the measured value of $r = 0.2$ ( tensor to scalar ratio), one can also estimate the energy scale for this accelerating regime to be around GUT scale ($10^{16} GeV$) which is below the Planck scale (the scale where the quantum gravity effects are prominent) but higher than the TeV scale ( the scale which can be probed by the current particle accelerator like LHC) (see \cite{ho} for an interesting discussion on this issue).

For inflationary model building using scalar fields, one needs to guarantee that there is sufficient slow-rolling for the scalar field to ensure necessary amount of inflation to solve the horizon and flatness problem. One also has to ensure the eventual breakdown of the slow-roll conditions so that the Universe exits from this inflationary phase and enters into a decelerated regime. This restricts  the shape of the potential for the inflaton.  Moreover most of the slow-roll scalar field models produce power-law type primordial power spectrum (PPS) of the form $P_{s,T} \sim A_{s,T} k^{n}$ where $A_{s,T}$ and $n$ are related to the the shape of the underlining potential for the inflaton. Measuring the temperature anisotropies in CMBR, one can put strong bounds on $A_{s,T}$ and $n$ and that puts further constraints on the inflaton potential (See \cite{martin} for scalar field models for inflation that are consistent with recent Planck results).

Recently Generalized Chaplygin Gas (GCG)  \cite{gcg} described by an equation of state $p = -\frac{A}{\rho^{\alpha}}$ where $A$ and $\alpha$ are constants, has been discussed widely in cosmological contexts. The case $(1+\alpha) > 0$ is interesting in the context of late time acceleration.  In this case, the GCG mimics dust in the early time and a dark energy with negative equation of state in late time. Initially considered as a natural candidate for a unification of dark matter and dark energy (UDM), it was later shown that this particular UDM behaviour is not suitable for the structure formation of the Universe \cite{udm}. But this fluid can be a possible dark energy candidate which tracks the background fluid initially and subsequently exits the tracking regime and starts acting as a dark energy candidate \cite{medhiraj}.

The opposite regime, $(1+\alpha) < 0$ is also equally interesting \cite{ss}. In this case, the GCG behaves like a cosmological constant ($w=-1$) initially but with time, the equation of state increases and becomes dust like ($w=0$). This behaviour is suitable for inflation, as in this case inflation happens initially and then automatically ends at later times.
We shall show that by properly adjusting parameters, one can get enough inflation that is necessary to solve horizon and flatness problems.

Motivated by this, we consider GCG as a model for inflation. We write the corresponding scalar field theory that mimics such behaviour and then calculate the PPS in this model. We show that in the context of Einstein gravity, to get the right shape for the PPS, the e-folding at the time of horizon exit ($N_{*}$) has to be excessively large which is a serious drawback. But if one considers a five dimensional brane-world scenario which results in a correction term in the Einstein equation, this problem of high e-folding at  horizon exit gets resolved and one gets a suitable inflationary model.

Recently, it has been pointed out that a power law form for the scalar PPS for the inflaton field is in tension with combined Planck+BICEP2 results \cite{hazra}. This is related to the fact that a significantly higher value for $r$  ($r=0.2$) as measured by BICEP2 is not consistent with the suppression of power in $C_{l}^{TT}$  at large scales as observed by Planck. In fact the authors in \cite{hazra} have shown that power law form for the scalar PPS with a single spectral index is ruled out at more than $3\sigma$ by Planck+BICEP2 in comparison to a broken PPS model containing two spectral indices (see \cite{solve} for different approaches to solve this problem). This is a bad news for inflationary model building because most of the standard and theoretically motivated slow-roll inflationary scenarios produce a power-law type scalar PPS. GCG as a inflationary model is of slow-roll type and also produces a power-law type scalar PPS. Hence this tension applies to GCG as well. But the underlining assumption for all these studies is that our Universe is described by a concordance $\Lambda$CDM model. We try to address this issue by going beyond the concordance $\Lambda$CDM model. By allowing a general dark energy equation of state, our study shows that one may address this issue even if one sticks to a power law type scalar PPS.

The structure of the paper is as follows: in section 2, we describe the GCG inflationary models and its scalar field representation; in section 3, we study the slow-roll inflationary models with GCG in Einstein gravity and discuss its problem; in section 4, we study the GCG inflation in a particular higher dimensional brane-world set up and put constraints on various model parameters using observational results from Planck and BICEP2 for a $\Lambda$CDM;  in section 5, we discuss the issue regarding the inconsistency between Planck and BICEP2 results with power-law type scalar PPS and try to address the issue with a general dark energy model. We do the full MCMC analysis with Planck+WP and BICECP2 data using a general dark energy model to get the constraint on our inflationary GCG model parameters and compare the results obtained using a $\Lambda$CDM model for dark energy; finally in section 6, we put our concluding remarks.

\section{The GCG Inflation}

The Generalized Chaplygin Gas (GCG) is described by the equation of state \cite{gcg}

\begin{equation}
p=-\dfrac{A}{\rho^{\alpha}},
\end{equation}

\noindent
where $p$ is the pressure and $\rho$ is the energy density of the GCG fluid. $A$ and $\alpha$  are the parameters of the model. One can calculate energy density as a function of the scale factor by integrating out the energy conservation equation in FRW background and  the corresponding expression is 

\begin{equation}
\rho (a) =(A+Ba^{m})^{-3/m},
\end{equation}

\noindent
where, $ m=-3(1+\alpha) $ and $B$ is an integration constant. For $ m < 0 $, GCG behaves like
a dust at very early time and it behaves like cosmological constant at infinite future. In between the equation of state smoothly changes over from dust behaviour to the cosmological constant. This behaviour is attractive for dark energy model building as the GCG tracks the background matter in the early time and then enters the dark energy regime in the late time. This is similar to tracking model of dark energy that attempts to solve the cosmic coincidence problem. On the other hand, this behaviour is not at all suitable for early time inflation, because in this case once GCG starts accelerating the Universe, it can never be stopped unless one invokes some extra effect to exit from the inflationary period (see \cite{infgcg} for inflationary model with GCG with $m <0$).

For $ m > 0 $ , the opposite happens. In this case the GCG behaves like a cosmological constant (CC) to start with, and then it slowly evolves away from this CC behaviour and eventually behaves like a dust. In this case, we have a inflationary epoch for early time which ends subsequently and the Universe enters into a decelerating dust like era. The end of inflation is automatic in this case without any need for extra mechanism. One can suitably choose the model parameters to get the required number of e-folds. With this, it is now important to see whether the primordial fluctuations that can be produced in such a model, is consistent with the observational results from experiments like Planck and BICEP2.

GCG can be described by a minimally coupled scalar field Lagrangian with a canonical kinetic energy term. The energy density and pressure for a canonical scalar field $\phi (t)$ which is minimally coupled to the gravity are given by (assuming flat FRW metric):

\begin{eqnarray}
\rho = \dfrac{1}{2}\dot{\phi}^{2}+V(\phi)\nonumber\\
p = \dfrac{1}{2}\dot{\phi}^{2}-V(\phi),
\end{eqnarray}

\noindent
where $V(\phi)$ is the potential for the field. Using these expressions and equations (1) and (2), together with the Einstein's equation 

\begin{equation}
3H^2 = \frac{1}{M_{Pl}^2} \rho,
\end{equation}

\noindent
one can write
\begin{equation}
\dfrac{d\phi}{da}=\sqrt{3}M_{Pl}\dfrac{1}{a\sqrt{1+ca^{-m}}},
\end{equation}

\noindent
where $ c = \dfrac{A}{B}$.  On integration, this results

\begin{equation}
A+Ba^{m}=A[\cosh(\dfrac{q(\phi -\phi_{0})}{2})]^{2},
\end{equation}

\noindent
where, $ q=-\dfrac{m}{\sqrt{3}M_{Pl}} $ and $ \phi_{0} $ is an integration constant. After some straightforward algebraic calculations, one can finally write

\begin{eqnarray}
V &=& V_{0} [\cosh(\dfrac{q(\phi -\phi_{0})}{2})]^{6\alpha/m} [3+\cosh(q(\phi -\phi_{0}))]\nonumber\\
H &=& H_{0} {[\cosh(\dfrac{q(\phi -\phi_{0})}{2})]^{-3/m}}, 
\end{eqnarray}

\noindent
where $V_{0}=\dfrac{1}{4}A^{-3/m} $ and $ H_{0} =  \sqrt{\dfrac{4V_{0}}{3M_{Pl}^{2}}} $.

\section{Slow-Roll Inflationary Model with GCG}

Using the expression for the Hubble parameter given in the previous section, one can now calculate the two Hubble slow-roll parameters $\epsilon_{H}$ and $\delta_{H}$ (one can also use the potential slow-roll parameters $\epsilon_{\phi}$ and $\eta_{\phi}$ and the results will be exactly the same). They are expressed as:

\begin{eqnarray}
\epsilon_{H}&&=2 M_{Pl}^{2} (\dfrac{H_{\phi}}{H})^{2} \nonumber\\
&&=\dfrac{3}{2} [\tanh(\dfrac{q(\phi -\phi_{0})}{2})]^{2}
\end{eqnarray}
\begin{eqnarray}
\delta_{H}&&=\epsilon_{H} -(\dfrac{\dot{\epsilon_{H}}}{2 H \epsilon_{H}}) \nonumber\\
&&=\dfrac{3}{2} -\dfrac{m+3}{2} [{\rm sech}(\dfrac{q(\phi -\phi_{0})}{2})]^{2}.
\end{eqnarray}

Under the slow roll approximation, the scalar spectral index $ n_{s}$ 
and tensor spectral index $ n_{T} $ are given by

\begin{eqnarray}
n_{s}&\simeq& (1-4\epsilon_{H} +2\delta_{H})\nonumber\\
n_{T}&\simeq& -2\epsilon_{H}.
\end{eqnarray}

Using the above expressions for $\epsilon_{H}$ and $\delta_{H}$, one can write their expressions as \cite{pert}:,

\begin{eqnarray}
n_{s}&\simeq& -2 +(3-m)[{\rm sech}(\dfrac{q}{2}(\phi -\phi_{0}))]^{2}\nonumber\\
n_{T}&\simeq& -3[\tanh(\dfrac{q}{2}(\phi -\phi_{0}))]^{2}.
\end{eqnarray}

The tensor to scalar ratio  $r$ which measures the amount of stochastic gravitational wave that is produced during inflation, is given by

\begin{equation}
r \simeq 16\epsilon_{H} = 24 [\tanh(\dfrac{q}{2}(\phi -\phi_{0}))]^{2}.
\end{equation}

One should note that all the relevant quantities like $n_{s}$, $n_{T}$, $r$ should be calculated at $aH=k$, i.e when a given mode of fluctuation exits the horizon. Once the fluctuations exit the horizon, they do not evolve and are frozen at their values  at horizon crossing. As the inflation ends, the horizon scale starts growing, and different fluctuations start entering inside the horizon. The fluctuations which are larger in scales enters later and fluctuations which are smaller in scales enter earlier. The fluctuations which are of horizon size today were the last to exit the horizon during inflation. Any scale $k_{*}$ today that we are interested in, can be related to the number of e-folds $N_{*}$ before the end of inflation which is given by \cite{pert}

\begin{equation}
N_{*} = \int^{t_{e}}_{t_{*}} H dt \simeq - \frac{1}{M_{Pl}^2} \int^{\phi_{e}}_{\phi_{*}} \frac{V}{V_{\phi}} d\phi,
\end{equation} 

\noindent
where subscript "e" denotes the end of inflation and subscript $\phi$ denotes differentiation w.r.t $\phi$.  For all the relevant scales that can be probed through CMB observations like Planck, $50<N_{*}<60$ \cite{plinf}.

Inflation will end when $ \epsilon_{H}=1 $. Using the expression (8), this gives

\begin{equation}
\phi_{e}=\phi_{0}-\dfrac{2}{q} \tanh^{-1}(\sqrt{2/3}) 
\end{equation}

The scale factor at the end of inflation can also be calculated from equation (6) as

\begin{equation}
a_{e}=(2c)^{1/m}.
\end{equation}
 
\noindent
Knowing that $N_{*} =\ln[a_{e}/a_{*}]$, one can now calculate the value of the scalar field at horizon exit, using equations (6), (14) and (15):

\begin{equation}
\phi_{H.E}=\phi_{0} -\dfrac{2}{q} [{\rm cosech}^{-1}(\sqrt{\dfrac{e^{m N_{*}}}{2}})].
\end{equation}

\noindent
Using this expression for $\phi_{H.E}$, , we get the value of the scalar spectral index and tensor-to-scalar ratio at horizon exit as:

\begin{eqnarray}
n_{s} (k=aH)= 1-m+\dfrac{2(m-3)}{2+\exp(m N_{*})}\nonumber\\
r (k=aH)=\dfrac{48}{2+\exp(m N_{*})}
\end{eqnarray}

%%%%%%%%%%%%%%%%%%%%%%%%%%%%%%%%%%%%%%%%
\begin{figure}
\includegraphics[scale=0.5]{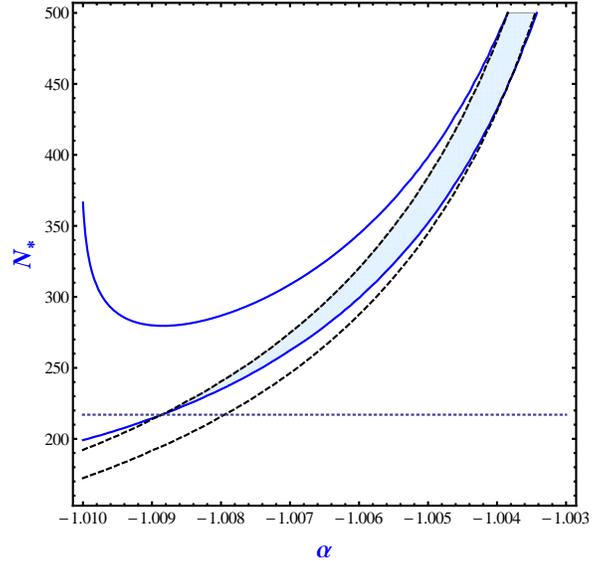}
\caption{Allowed region in the ($\alpha-N_{*}$) plane. The region inside the solid lines are for $n_{s}$ constraint from Planck+WP, the region inside the dashed lines are for $r$ constraint from BICEP2. The constraints are for $\Lambda$CDM model. The shaded region satisfies both the constraints.}
\label{fig:chart2}
\end{figure}
%%%%%%%%%%%%%%%%%%%%%%%%%%%%%%%%%%%%%%%%%%%%%

The current bound on $ n_{s} $  and $r$ as obtained by Planck+WP data \cite{planck} and  BICEP2 \cite{bicep2} are the following:

\begin{eqnarray}
n_{s} &=& 0.9624\pm 0.0075\nonumber\\
r &=& 0.20^{+0.07}_{-0.05}.
\end{eqnarray}

 We should mention that these bounds are obtained using a $\Lambda$CDM dark energy model. In section 5, we obtain the similar bound assuming a more general dark energy model.

In Figure (1), we draw the contours in the $(\alpha, N_{*})$ plane which satisfy the above constraints. One can see that minimum value for $N_{*}$ required is $N_{*} \simeq 217$ which is way above the theoretical prior $50<N_{*}<60$. This is the main drawback of this model. 
 
So, the generalized chaplygin gas is not a suitable model for inflation in Einstein's gravity.

\section{Inflationary Model in Brane-World Scenario with GCG}

In this scenario, we consider the observable Universe to be confined on a 3-brane embedded in a 5-D anti de-Sitter spacetime. One such scenario was first proposed in 1998 by Randall and Sundrum (RS) \cite{rs} to solve the hierarchy problem in particle physics. This scenario consists of a 5-D space-time governed by Einstein gravity with a negative cosmological constant in the bulk. The space-time respects the $S^{1}/Z_{2}$ symmetry and the flat 3-branes are located at orbifold fixed points in this geometry. One of the branes is our visible Universe where the modified Einstein equation is now given by \cite{binet}:

\begin{equation}
H^{2} = \frac{1}{3 M_{Pl}^2} \rho\left(1+\frac{\rho}{2\lambda}\right),
\end{equation}

\noindent
 where, $ \lambda $ is the 3-brane tension.  The relation between $ \lambda $ and five-dimensional Planck mass $ M_{5} $ is given by

\begin{equation}
M_{5}=(\dfrac{4 \pi \lambda}{3})^{1/6}  M_{Pl}^{1/3}.
\end{equation}

\noindent
In this scenario, the potential slow-roll parameters are given by \cite{mar1,bent}

\begin{eqnarray}
\epsilon_{\phi} &=&\dfrac{1}{2} M_{Pl}^{2} (\dfrac{V_{\phi}}{V})^{2} \dfrac{1+\dfrac{V}{\lambda}}{(1+\dfrac{V}{2\lambda})^{2}}\nonumber\\
\eta_{\phi} &=& M_{Pl}^{2}\dfrac{V_{\phi \phi}}{V}\dfrac{1}{1+\dfrac{V}{2\lambda}}
\end{eqnarray}

\noindent
In the high energy limit  $\dfrac{V}{\lambda} >> 1 $, one can approximate  the slow-roll parameters as

\begin{equation}
\epsilon_{\phi} =2 M_{Pl}^{2} (\dfrac{V_{\phi}}{V})^{2}\dfrac{\beta}{\bar{V}},
\end{equation}
\begin{equation}
\eta_{\phi} =2M_{Pl}^{2}\dfrac{V_{\phi \phi}}{V}\dfrac{\beta}{\bar{V}},
\end{equation}

\noindent
where $\bar{V} = \frac{V}{V_{0}}$ and $ \beta =\dfrac{\lambda}{V_{0}} $. The scalar spectral index $ n_{s} $ can be written in terms of the  potential slow-roll parameters as:

\begin{equation}
n_{s}=1-6\epsilon_{\phi} +2\eta_{\phi}
\end{equation}

\noindent
The amplitude of the scalar as well as the tensor perturbations are defined as \cite{mar2}

\begin{equation}
A_{s}^{2}\simeq \dfrac{1}{12\pi^{2}M_{Pl}^{6}} \dfrac{V^{3}}{V_{\phi}^{2}} [1+\dfrac{V}{2\lambda}]^{3}
\end{equation} 

\begin{equation}
A_{t}^{2} \simeq \dfrac{2}{3\pi^{2}M_{Pl}^{4}} V (1+\dfrac{V}{2\lambda})F^2
\end{equation}

\noindent
where, 
\begin{equation}
F^{2}=[\sqrt{1+s^{2}}-s^{2} \sinh^{-1}(\dfrac{1}{s})]^{-1}
\end{equation} 
\begin{equation}
s=[\dfrac{2V}{\lambda}(1+\dfrac{V}{2\lambda})]^{1/2}
\end{equation}

\noindent
Using Equations (25) and (26), we can now write the tensor-to-scalar ratio $r$ as,

\begin{equation}
r = \dfrac{A_{t}^{2}}{A_{S}^{2}}=8 M_{Pl}^{2} (\dfrac{V_{\phi}}{V})^{2} [1+\dfrac{V}{2\lambda}]^{-2} F^{2}
\end{equation}

\noindent
In the high energy limit $\frac{V}{\lambda} >> 1$, $ F^{2}=\dfrac{3V}{2\lambda} $ and  hence in this limit, $ r $ becomes

\begin{equation}
r=24\epsilon_{\phi}
\end{equation}

\noindent
Using the above definitions, we get the expressions for the slow-roll parameters for our model as

\begin{eqnarray}
\epsilon_{\phi} =&&\dfrac{2}{3}\beta [\cosh(\dfrac{q}{2}(\phi -\phi_{0}))]^{6/m}\nonumber\\
&&\times[9+2m+3\cosh(q(\phi -\phi_{0}))]^{2} \nonumber \\
&&\times[\sinh(\dfrac{q}{2}(\phi -\phi_{0}))]^{2}\nonumber\\
&&\times[3+\cosh(q(\phi -\phi_{0}))]^{-3}
\end{eqnarray}

\begin{eqnarray}
\eta_{\phi} =&&\dfrac{\beta}{6} [\cosh(\dfrac{q}{2}(\phi -\phi_{0}))]^{6/m}\nonumber\\
&&\times[3+\cosh(q(\phi -\phi_{0}))]^{-2} \nonumber\\
&&\times[9\cosh(2q(\phi -\phi_{0}))\nonumber\\
&&+2(2m^{2}+9m+18)\cosh(q(\phi -\phi_{0})) \nonumber\\
&&-(4m+15)(2m+3)]
\end{eqnarray}

\noindent
We also calculate the value of the field $ \phi $ at horizon exist as

\begin{equation}
\phi_{H.E}=\phi_{0}-\dfrac{2}{q}[{\rm cosech}^{-1}(\sqrt{2e^{m N_{*}}})]
\end{equation} 

\noindent
where $N_{*}$ is the number of e-fold at the horizon exit.  Using this, we calculate different observables like $n_{s}$, $r$ at the horizon exit as

\begin{eqnarray}
n_{s} &&=1 - \dfrac{2\beta}{3} [1+\dfrac{1}{2}e^{-mN_{*}}]^{3/m} [1+4e^{mN_{*}}]^{-3} 
\nonumber\\
&&\times[18+(144+27m-2m^{2})e^{mN_{*}}\nonumber\\
&& +6(48+20m+m^2)e^{2mN_{*}}\nonumber\\
&&+8m(m+6)e^{3mN_{*}}]
\end{eqnarray}
%%%%%%%%%%%%%%%%%%%%%%%%%%%%%%%%%%%%

\begin{figure*}[t]
    \centering
    \subfigure
    {
        \includegraphics[scale=0.35]{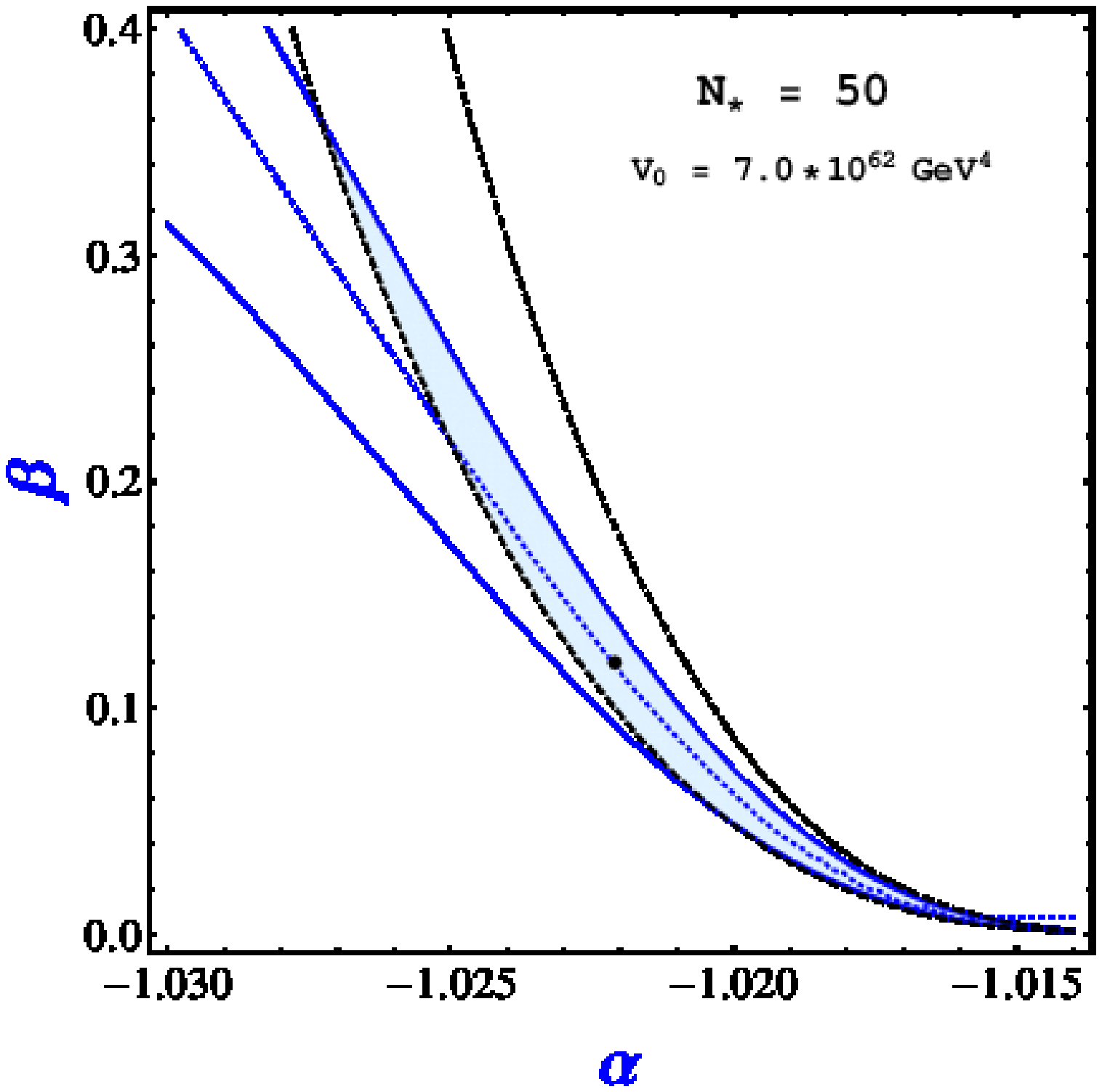}
    }
    \\
    \subfigure
    {
        \includegraphics[scale=0.35]{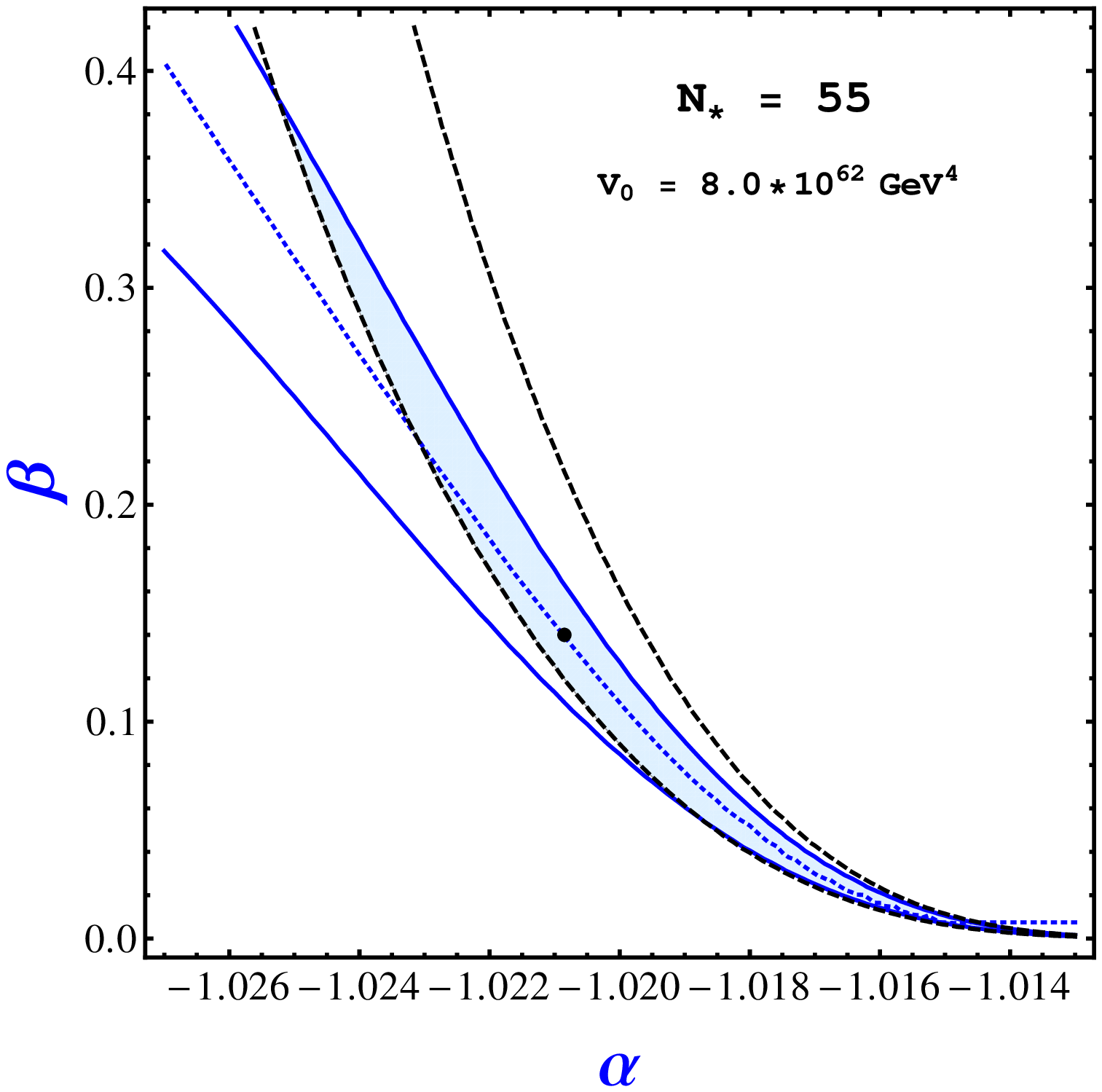}
    }
    \subfigure
    {
        \includegraphics[scale=0.36]{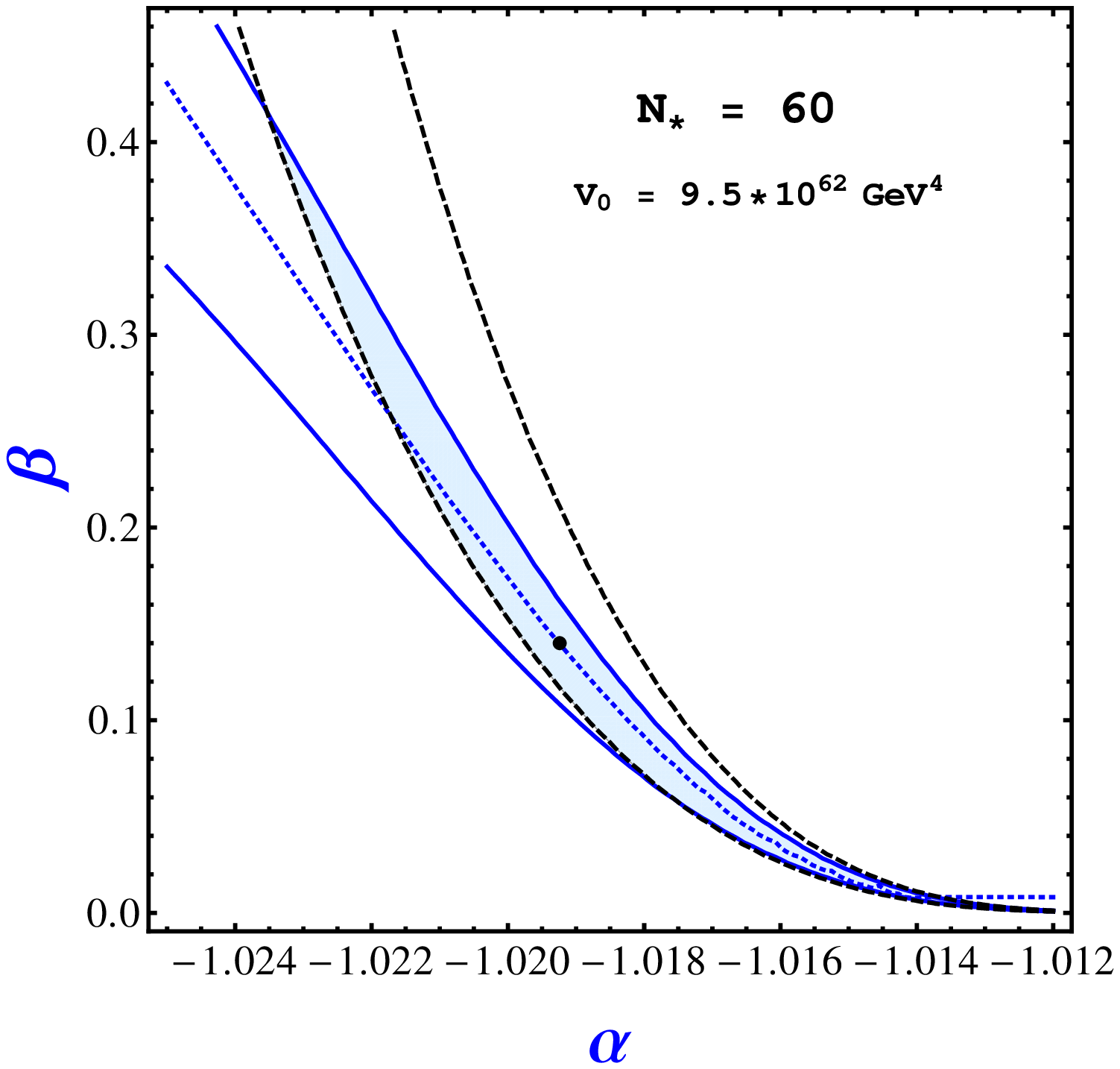}
    }
    \caption{Allowed region in ($\alpha-\beta$) parameter space for different values of $N_*$. Here we have used best fit values and bounds of $n_s$ and $r$ given by Planck+WP+BICEP2. The cosmological model is assumed to $\Lambda$CDM. The region inside the solid lines satisfies the constraint on $n_{s}$ from Planck+WP and the region inside the dashed lines satisfies the constraint on $r$ from BICEP2. The shaded region satisfies both the constraints. The dotted lines satisfies the constraint on $A_{s}$ from Planck for $V_{0} = 7.0 \times 10^{62}$ $GeV^{4}$(for top figure), $8.0 \times 10^{62}$ $GeV^{4}$(bottom left) and $9.5 \times 10^{62}$ $GeV^{4}$(bottom right). The dots are for the $M_{5} = 1.47\times 10^{-2} M_{Pl}$(top figure), $M_{5} = 1.54\times 10^{-2} M_{Pl}$ (bottom left) and $M_{5} = 1.58\times 10^{-2} M_{Pl}$ (bottom right).}
\end{figure*}

%%%%%%%%%%%%%%%%%%%%%%%%%%%%%%%%%%

\begin{equation}
r=8\beta [1+\dfrac{1}{2}e^{-mN_{*}}]^{3/m} [1+4e^{mN_{*}}]^{-3} 
[3+2(6+m)e^{mN_{*}}]^{2}
\end{equation}
 
 \noindent
The current bound on these two parameters from Planck+WP+BICEP2 with a $\Lambda$CDM model are given in equation (18). Next, we calculate the initial value of the field  $ \phi  $ which gives total 70 e-folds of inflation, necessary to solve the horizon problem:
 
\begin{equation}
\phi_{i}=\phi_{0}-\dfrac{2}{q}[{\rm cosech}^{-1}(\sqrt{2e^{m N_{total}}})]
\end{equation}

\noindent
where, $ N_{total} $ is total number of e-folds during inflation. From equation (25) and using the high energy limit, we get the expression for $ A_{s}^{2} $ at the horizon exit as

\begin{eqnarray}
A_{s_{H.E}}^{2} &&=\dfrac{V_{0}}{2\pi^{2} \beta^{3}M_{Pl}^{4}} [1+\dfrac{1}{2}e^{-m N_{*}}]^{- 12/m} \nonumber\\
&&\times[1+4e^{m N_{*}}]^{6} [1+2e^{m N_{*}}]^{-3}\nonumber\\
&&\times[2(m+6)e^{m N_{*}} +3]^{-2}
\end{eqnarray}

\noindent
The measured value for $A_{s} ^{2}$ by Planck for $\Lambda$CDM model is given as $\ln(10^{10} A_{s}^{2}) = 3.089$ \cite{planck}. Subsequently we study the parameter space ($\alpha,\beta$) that is allowed by the results obtained by Planck+WP and BICEP2 for the observables $n_{s}$, $r$ and $A_{s}$. This is shown in figure (2) for different values of $N_{*}$.  We should again stress that these results are obtained assuming a $\Lambda$CDM Universe.

The first thing to be noticed is that we now have a parameter space that is allowed by the Planck+WP+BICEP2 data with $50\leq N_{*}\leq 60$. This is due to the modified Einstein's equation in RS brane-world set up.

%%%%%%%%%%%%%%%%%%%%%%%%%%%%%%%
\begin{center}
\begin{figure*}[t]
\begin{tabular}{c@{\quad}c}
\epsfig{file=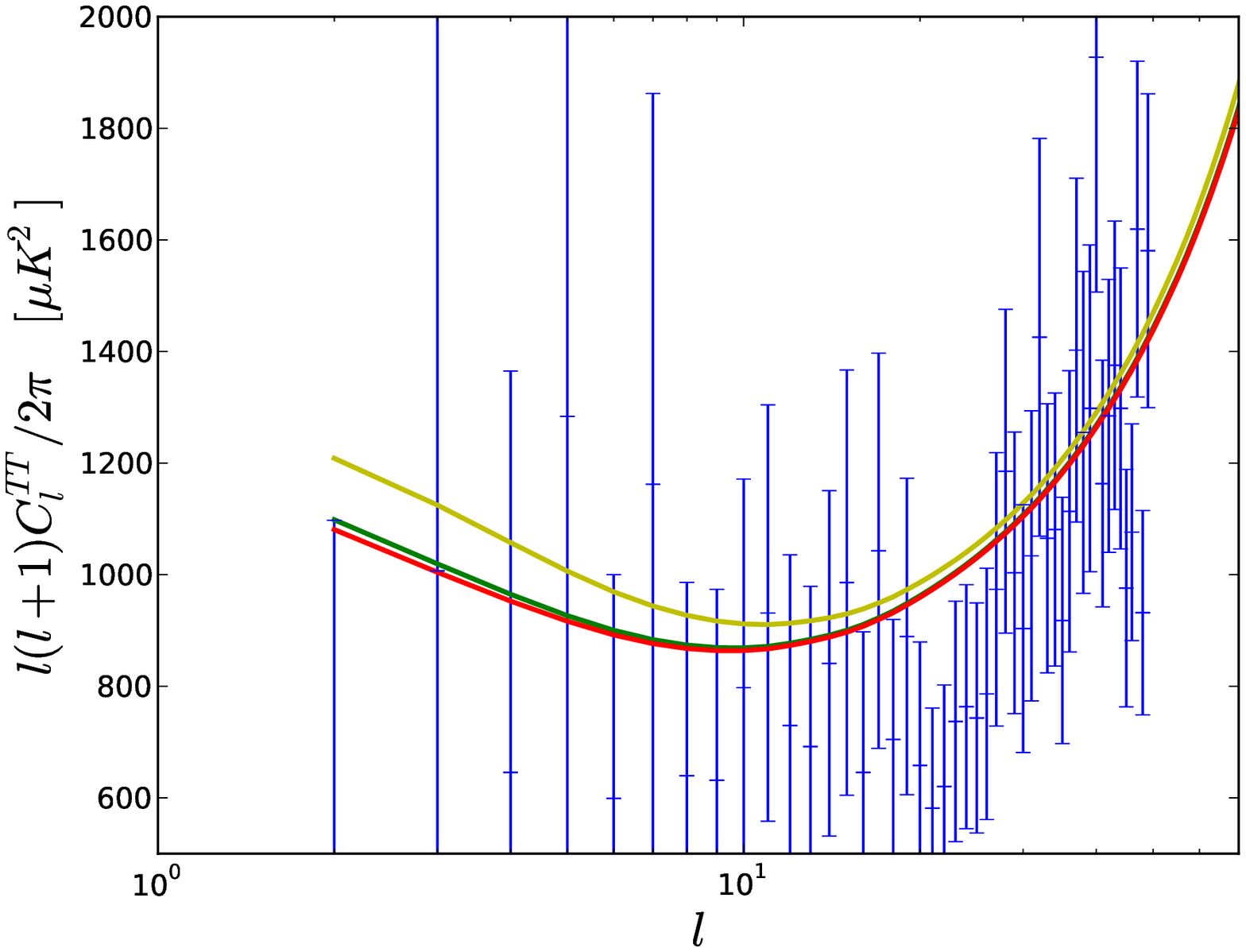,width=7 cm}&
\epsfig{file=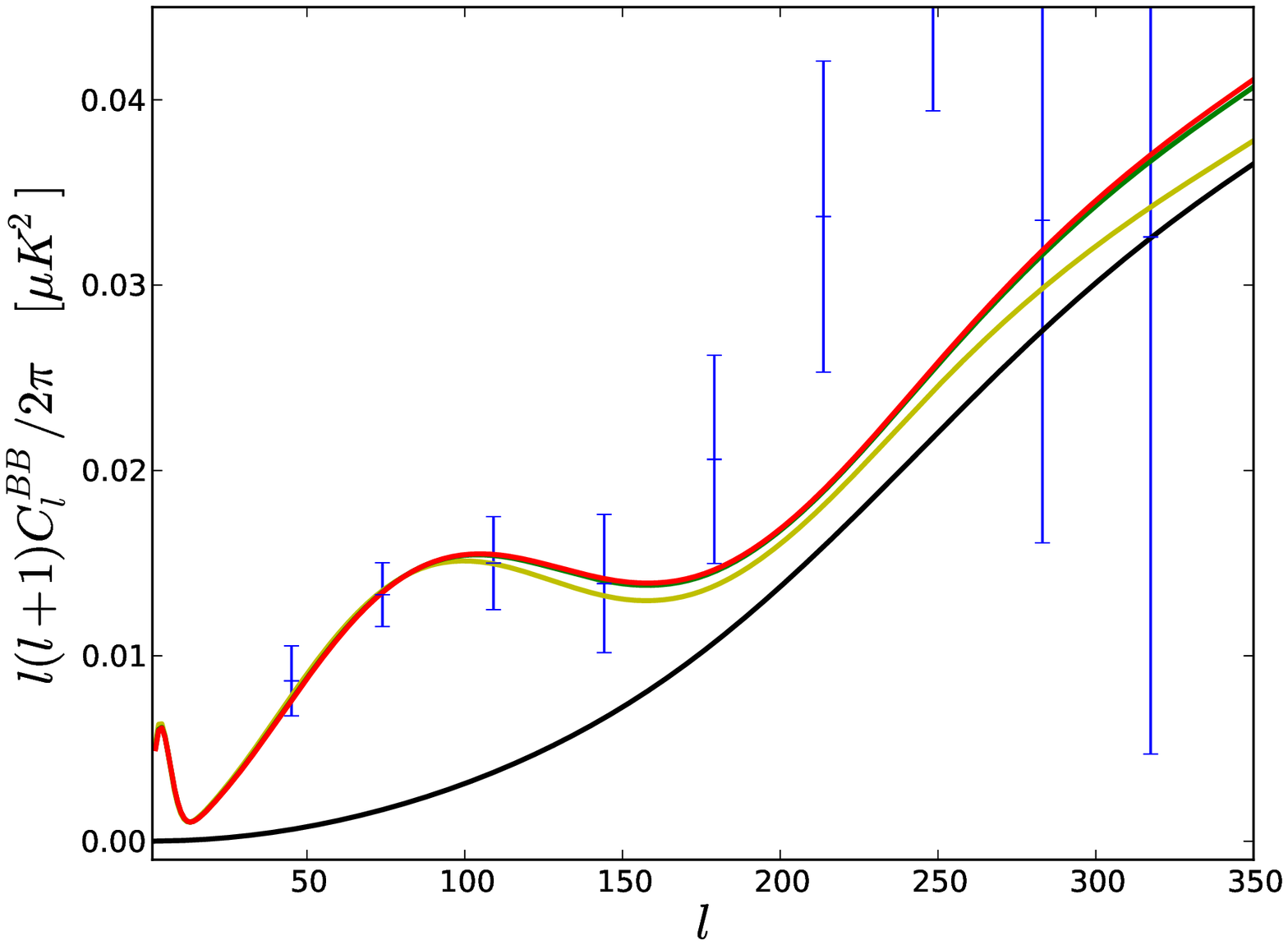,width=7 cm}\\
\end{tabular}
\caption{Behaviour of $C_{l}^{TT}$ (left) and $C_{l}^{BB}$ (right) with angular scale $l$ for different cases. For left figure, top to bottom, Power Law PPS ($\Lambda$CDM)+ r (Planck+WP+BICEP2), Power Law PPS (DE1)+ r (Planck+WP+BICEP2), Power Law PPS (DE2)+ r (Planck+WP+BICEP2). The error bars are for Planck low-l data points. For right figure, bottom to top,  Power Law PPS ($\Lambda$CDM)+ r (Planck+WP), Power Law PPS ($\Lambda$CDM)+ r (Planck+WP+BICEP2), Power Law PPS (DE1)+ r (Planck+WP+BICEP2), Power Law PPS (DE2)+ r (Planck+WP+BICEP2). The error bars are for BICEP2 data points.}
\end{figure*}
\end{center}

%%%%%%%%%%%%%%%%%%%%%%%%%%%%%%%%%

In these figures, the shaded regions represent the allowed parameter space that satisfies the constraints on $n_{s}$ and $r$ as obtained by Planck and BICEP2 respectively.  Once we have such a allowed region, one can then fix the energy scale of the inflation as given by $V_{0}$ which satisfies the constraint on $A_{s}$. The dotted line in each figure represents a typical behaviour in ($\alpha,\beta$) parameter space for a particular choice of $V_{0}$. The value of $V_{0}$ is chosen in such a way so that this line falls within the shaded region. This shows that typical scale of inflation is around $5\times 10^{15}$ GeV, just below the GUT scale. We should emphasis that the constraint on $A_{s}$ is very sensitive on this scale as slight deviation from this scale can move the dotted line outside the shaded region making it inconsistent with the allowed region for $n_{s}$ and $r$. As $V_{0}$ is related to the brane tension $\lambda$ which in turn is related to the five dimensional Planck mass $M_{5}$, one can also estimate the 5-D Planck mass $M_{5}$. These are $ M_{5} = 1.47 \times 10^{-2} M_{Pl} $, $ M_{5} = 1.54 \times 10^{-2} M_{Pl} $ and $ M_{5} = 1.58 \times 10^{-2} M_{Pl} $ for $N_{*} = 50,55,60$ respectively. From eqn (36), one can also calculate $\phi_{i}$ necessary to have $70$ e-folds of total inflation. The required value for $\phi_{i}$ typically varies from $1.6 M_{Pl}$ to $19.2 M_{Pl}$ for different $N_{*}$.

\section{Constraints with CPL dark energy}

As discussed in the Introduction, it is now known that a measured high value for $r$ by BICEP2 is in tension with suppression of power at large scale as observed by Planck. In fact the recent paper by Hazra et al. \cite{hazra} has ruled out the simple power-law form for scalar PPS in comparison to a broken scalar  PPS using Planck+BICEP2 data at more than $3\sigma$. And this is precisely due to the fact that with a power-law form for the scalar PPS, a high value of $r$  ($r=0.2$) as measured by BICEP2 is inconsistent with the suppression of power at large angular scales as observed by Planck. But this suppression can be achieved with a broken scalar PPS as shown in \cite{hazra}. Their investigation assumes a concordance $\Lambda$CDM model for our Universe.

\renewcommand{\arraystretch}{1.1}

\begin{table*}[!htb]
\begin{center}
\vspace{4pt}
\begin{tabular}{|c | c | c || c | c |}
\hline\hline
 \multicolumn{5}{|c|}{{\bf Comparison of the $\Lambda$CDM with CPLDE}}\\

 \hline
 & \multicolumn{2}{|c||}{{\bf Planck + WP}}& \multicolumn{2}{|c|}{{\bf Planck + WP + BICEP2}}\\
\cline{2-5}

$n_{\rm T}=-r/8$& $\Lambda$CDM & CPLDE & $\Lambda$CDM & CPLDE \\

\hline

$\Omega_{\rm b}h^2$ & 0.02217 & 0.0223 & 0.0221 & 0.0223\\
\hline

$\Omega_{\rm CDM}h^2$ & 0.1183 &  0.1171&0.1177 &0.116\\
\hline

$100\theta$ & 1.041 & 1.042 &1.041 &1.041\\
\hline

$\tau$ & 0.088 & 0.088 & 0.089 & 0.089\\
\hline

$n_{s}$ & 0.9658 & 0.9676 & 0.9686 & 0.9732\\
\hline

$w_{0}$ & -1 & -1.408 & -1 & -1.599\\
\hline

$w_{a}$ & 0 & -0.894 & 0 & -1.17\\
\hline

$r$ & 0.009 & 0.01 & 0.16 & 0.17\\
\hline

${\rm{\ln}}(10^{10} A_{\rm S})$ & 3.085 & 3.081 & 3.085 & 3.081\\

\hline
\hline
\multicolumn{5}{c}{$-2\ln{\cal L}$ [Best fit]}\\
\hline
{\tt commander} & -7.454 & -8.61 &-1.695 &-4.802\\
\hline
{\tt CAMspec} & 7796.235 & 7795.474 & 7797.54&7796.988\\
\hline
{\tt WP} & 2014.141 & 2014.55 & 2013.321&2013.572\\
\hline
{\tt BICEP2} & - & - &39.141 &38.281\\
\hline
Total & 9802.92 & 9801.41 &9848.31 &9844.04\\
\hline
$-2\Delta\ln{\cal L}$ & - & -1.51 & - &-4.3\\
\hline

\hline\hline
\end{tabular}
\end{center}
\caption{Results for the $\chi^2$ minimization with COSMOMC.  For all cases, power-law form for the scalar PPS is assumed. }
\end{table*}

Here we keep the power-law form for the scalar PPS but we deviate from the concordance $\Lambda$CDM model by introducing a dynamical dark energy model given by the equation of state parameter as prescribed by Chevallier and Polarski \cite{chev} and Linder \cite{lind}:

\begin{equation}
w = w_{0} + w_{a} (1-a).
\end{equation}

In figure (3), we show the $C_{l}^{TT}$ and $C_{l}^{BB}$ variations with angular scale $l$ using the publicly available code CAMB \cite{camb}. The consistency relation $r=-8 n_{T}$ is assumed. We use the parameter initialization values as provided by BICEP2 \cite{bicepini}. We also fix $r=0.2$ for our purpose. The top-most yellow line represents the $C_{l}^{TT}$ behaviour with a power law PPS together with a concordance $\Lambda$CDM model. It is easy to see the large enhancement of power at large scales, specially at $l=2$, makes it inconsistent with the Planck measurements for $C_{l}^{TT}$. With a concordance $\Lambda$CDM model and a power-law type scalar PPS, it is hard to escape from this inconsistency.  

%%%%%%%%%%%%%%%%%%%%%%%%%%
\begin{center}
\begin{figure*}[!h]
\begin{tabular}{c@{\qquad}c}
\epsfig{file=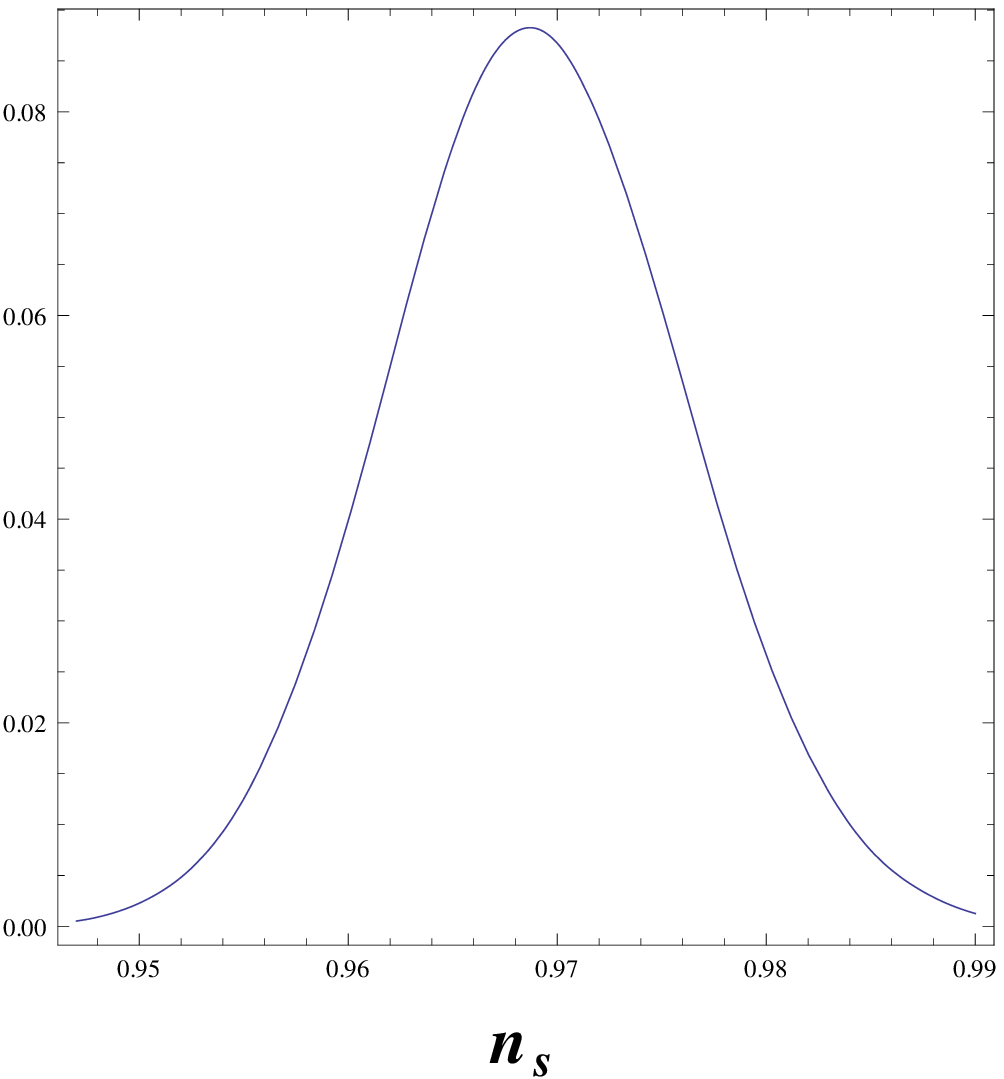,width=6 cm}&
\epsfig{file=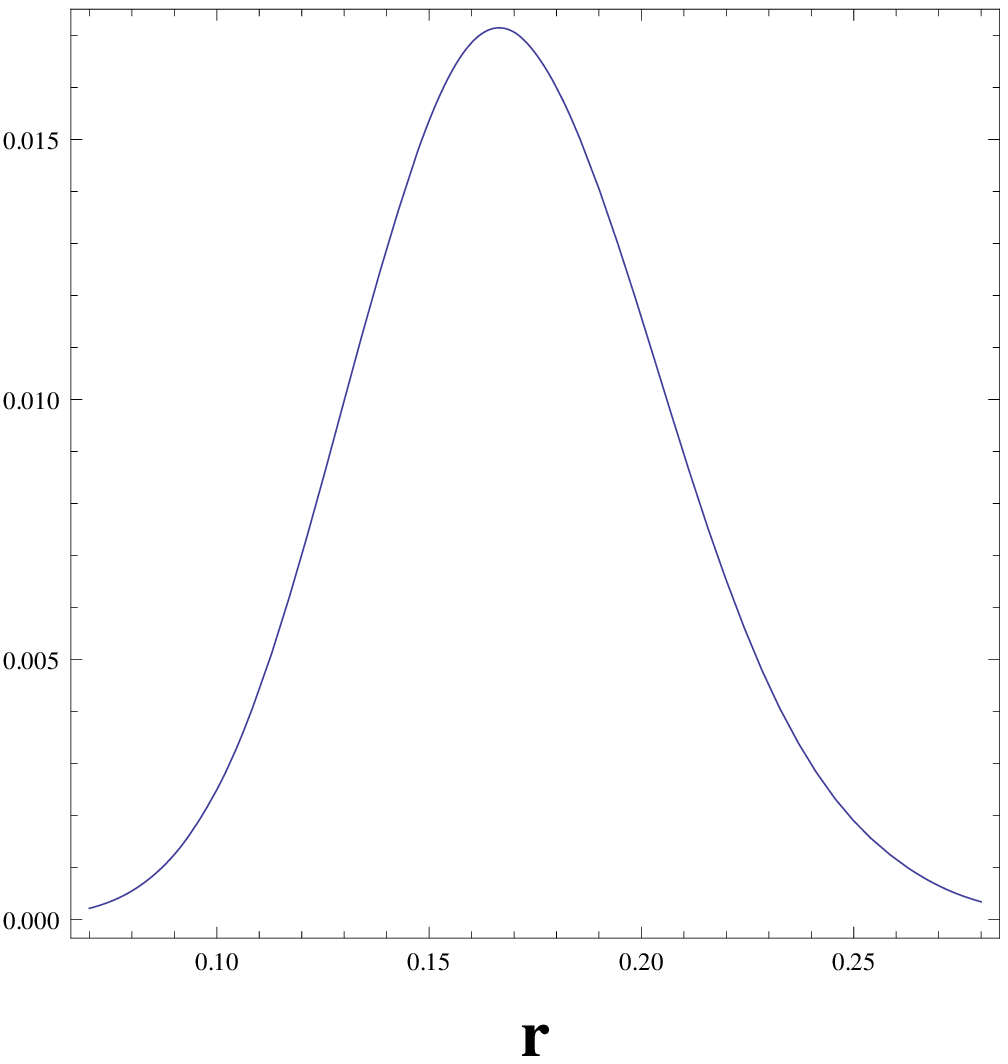,width=6 cm}\\
\end{tabular}
\caption{Likelihood functions for $n_s$ and $r$ using Planck+WP+BICEP2 with CPLDE model.}
\end{figure*}
\end{center}
%%%%%%%%%%%%%%%%%%%%%%%%%%

%%%%%%%%%%%%%%%%%%%%%%%%%%%%%%%%%%
\begin{center}
\begin{figure*}[!h]
\begin{tabular}{c@{\quad}c}
\epsfig{file=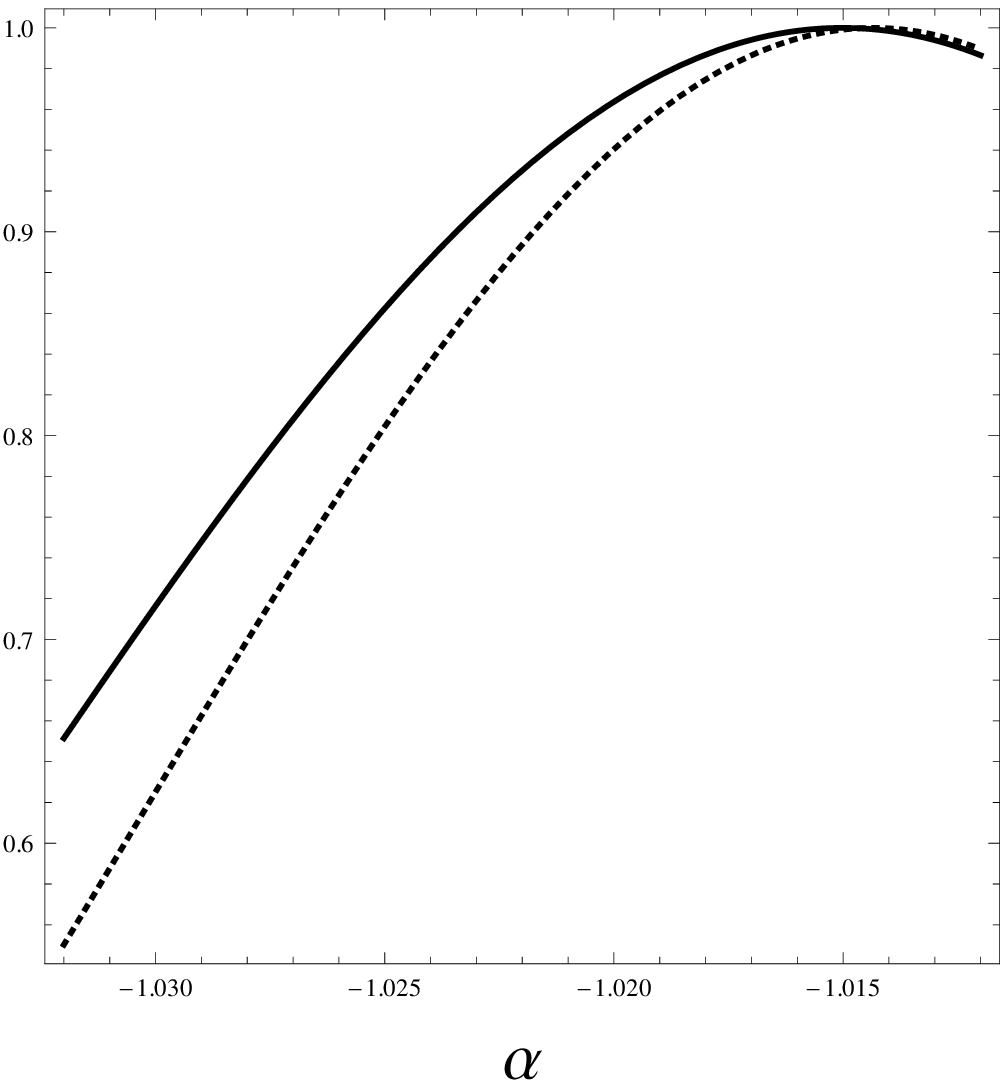,width=6 cm}&
\epsfig{file=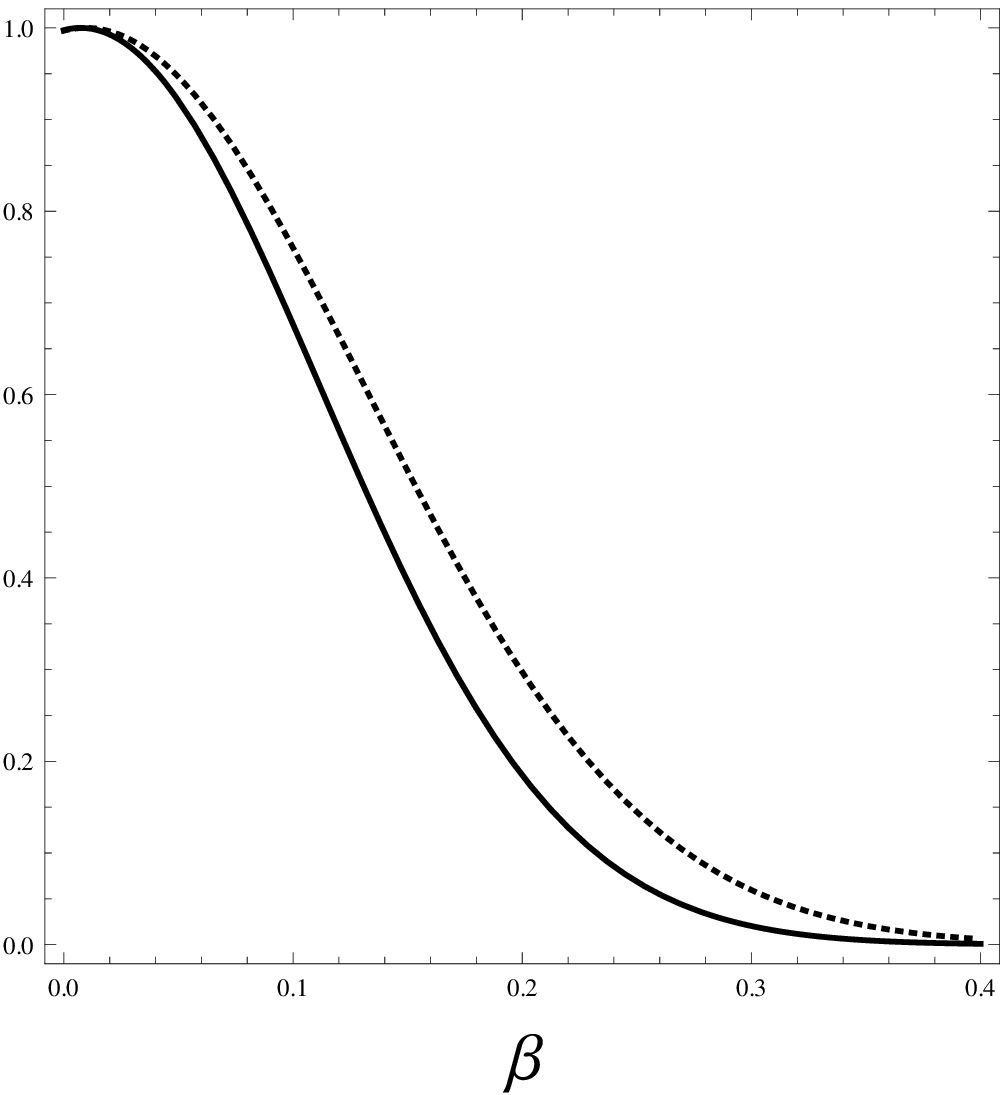,width=6 cm}\\
\end{tabular}
\caption{Likelihood function for $\alpha$ and $\beta$ for different values of $N_{*}$ using Planck+WP+BICEP2 with CPLDE model. Thick black lines in both figure represents $N_* = 55$ while dotted lines in both figures corresponds to likelihood for $N_* = 60$.}
\end{figure*}
\end{center}

%%%%%%%%%%%%%%%%%%%%%%%%%%%%%%%%%%%
To ease this tension, we extend the parameter space by allowing a phantom equation of state ($w_{0} <-1$) together with $w_{a} <0$ ( so that the equation of state remains phantom at all times) for the dark energy. To demonstrate how this may address the issue, we take two possible phantom dark energy models  (DE1 and DE2) with two specific choices for the parameters $w_{0}$ and $w_{a}$, e.g., $w_{0} = -1.3, w_{a} = -0.5$ and $w_{0} = -1.4, w_{a} = -0.5$ respectively.  The rest of the parameters are fixed as in the $\Lambda$CDM case described above. With these choices, one can now suppress the power at large scales to make it more consistent with Planck measurement as one can see from the plots in Figure (3). Moreover, the $C_{l}^{BB}$ plot shows that if one assumes $r=0.2$ with a power-law type scalar PPS, both of these phantom models  are consistent with B-mode polarization measurement by BICEP2. 

\begin{figure*}
    \centering
    \subfigure
    {
        \includegraphics[scale=0.35]{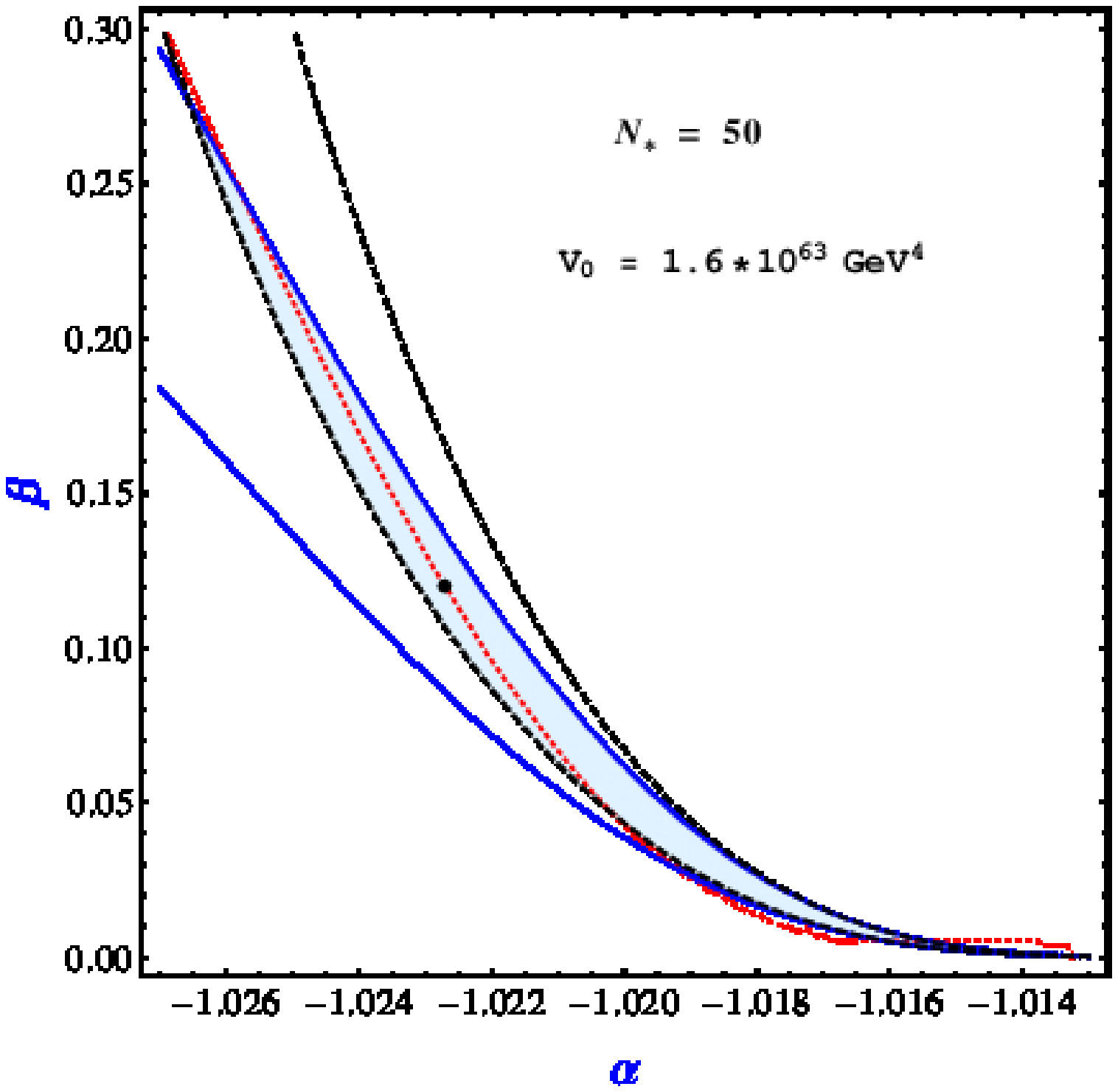}
    }
    \\
    \subfigure
    {
        \includegraphics[scale=0.35]{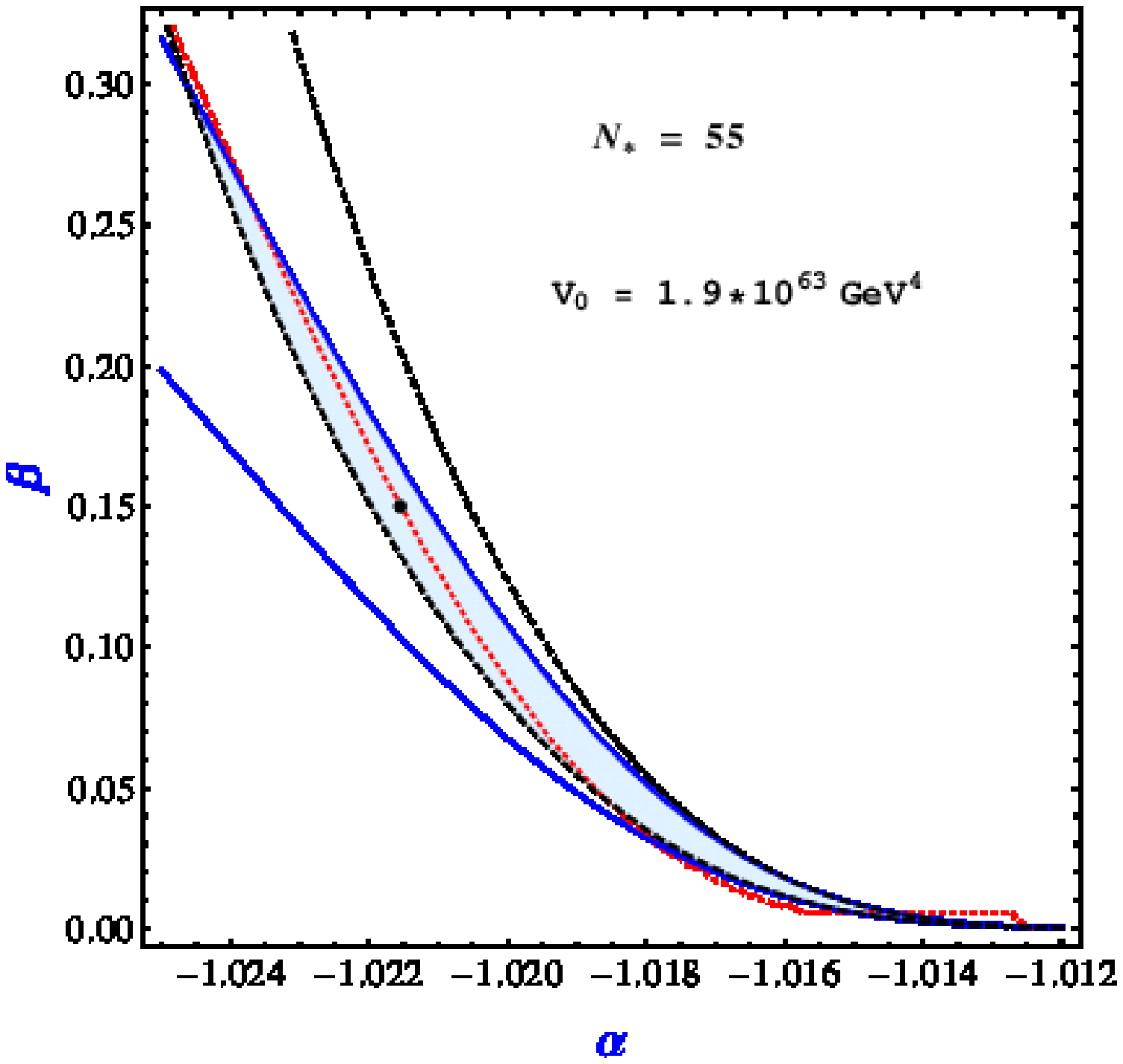}
    }
    \subfigure
    {
        \includegraphics[scale=0.36]{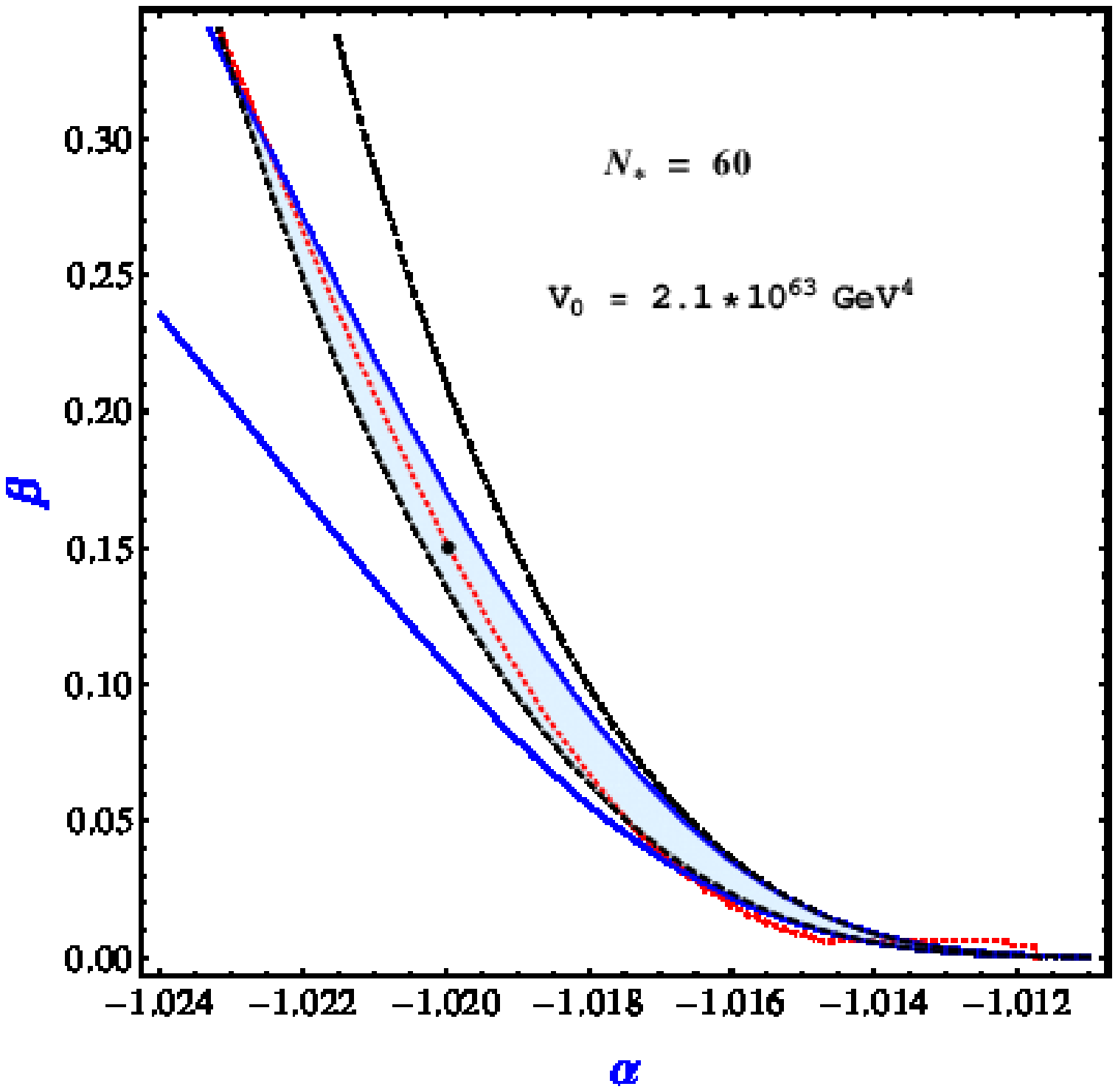}
    }
    \caption{Allowed region in ($\alpha-\beta$) parameter space for different values of $N_*$ using Planck+WP+BICEP2 with CPLDE model. The region inside the solid lines satisfies the constraint on $n_{s}$ and the region inside the dashed lines satisfies the constraint on $r$. The shaded region satisfies both the constraints. The dotted lines satisfies the constraint on $A_{s}$ from Planck for $V_{0} = 1.6 \times 10^{63}$ $GeV^{4}$(for top figure), $1.9 \times 10^{63}$ $GeV^{4}$(bottom left) and $2.1 \times 10^{63}$ $GeV^{4}$(bottom right). The dots are for the $M_{5} = 1.68\times 10^{-2} M_{Pl}$,(top figure), $M_{5} = 1.8\times 10^{-2} M_{Pl}$ (bottom left) and $M_{5} = 1.83\times 10^{-2} M_{Pl}$ (bottom right).}
\end{figure*}

%%%%%%%%%%%%%%%%%%%%%%%%%%%%%%%%%%%%%%%%%%
Next, we  also calculate the best fit $\chi^2$ values for $\Lambda$CDM and a CPL dark energy model (CPLDE) using the Planck+WP and BICEP2 likelihoods.  We use only the $\chi^2$ minimization routine in the publicly available code COSMOMC \cite{cosmomc} for this. The results are shown in Table I.

Without the BICEP2 data, for CPLDE models, there is an improvement around $1.51$ in $\chi^2$ compared to the $\Lambda$CDM model. This is consistent with the earlier results obtained by Hazra et al. \cite{medhiraj}. But with the addition of BICEP2 data, the improvement in $\chi^2$ is roughly $4.3$ for CPLDE models which is roughly three times of what one gets without BICEP2. We should stress that these numbers are indicative. It shows that with a full likelihood analysis for CMB+non-CMB data using MCMC, one can expect a substantially better fitting with a general dark energy keeping the power-law form for scalar PPS. In a recent paper, Hazra et al. \cite{medhiraj} have shown that there is a mild preference for phantom model over $\Lambda$CDM with the current CMB +non-CMB data (pre BICEP2) and $\Lambda$CDM is disfavoured at more than $1\sigma$ confidence level but it is still allowed at $2\sigma$. All these happen with simple power law type scalar PPS.  Our result shows that with the inclusion of BICEP2 data, this result can change substantially. 

bf Encouraged by this, we run the full MCMC chain with COSMOMC for a general CPLDE model taking the Planck+WP+BICEP2 data. As before, we use the parameter initialization values as provided by BICEP2 \cite{bicepini}. Additionally, we use the gaussian prior for $w_{0}$ and $w_{a}$ with central values as $-1.1, -0.5$ and the standard deviations $0.01, 0.1$ respectively .

The likelihoods for $n_{s}$ and $r$ are shown in figure 4. The $68.3\%$ error bars for $n_{s}$ and $r$ are 

\begin{eqnarray}
n_{s} = 0.9690\pm 0.0071\nonumber\\
r = 0.1707 \pm 0.0367
\end{eqnarray}

\noindent
One can see that going beyond concordance $\Lambda$CDM model, the allowed value of $r$ has come down appreciably and at $68.3\%$ confidence limit, it is almost same as what Planck+WP obtained for $r$, i.e $r<0.11$. 

Next, using the covariance between $n_{s}$ and $r$ as obtained from the full COSMOMC chains, we obtain the corresponding likelihood functions for GCG model parameters $\alpha$ and $\beta$ and subsequently the likelihood contours in the $\alpha-\beta$ parameter plane. These are shown in figures 5 and 6.  One can see from figure 6 that the allowed regions in $\alpha-\beta$ parameter space is decreased slightly compared to those obtained using a $\Lambda$CDM model as shown in figure 2. 

Finally, in figure 7,  we draw the contours in the $n_{s}-r$ parameter space using Planck+WP+BICEP2 with CPLDE model and show different combinations of $\alpha$ and $\beta$ which are allowed in this parameter space. Here we also show the same contours using only the Planck+WP data. It is evident that going beyond the concordance $\Lambda$CDM model, these two data sets are consistent. 

%%%%%%%%%%%%%%%%%%%%%%%%%%%%%%%%%%

%%%%%%%%%%%%%%%%%%%%%%%%%%%%%
\begin{figure}[!h]
\centering
\includegraphics[scale=0.5]{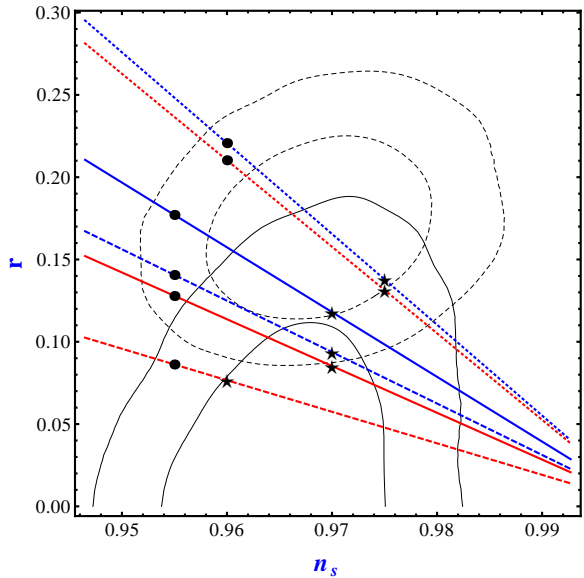}
\caption{$68\%$ and $95\%$ contour regions in $ r-n_{s} $ plane allowed by Planck+WP data (Solid Contours), and by Planck+WP+BICEP2 data (Dashed Contours)  using CPLDE. Dotted, continuous and dashed lines correspond to $ \alpha = -1.016 $, $ \alpha = -1.025 $ $ \& $ $ \alpha = -1.028 $ respectively. For each of these cases, the upper line is for $N_{*} = 50$ and the lower one is for $N_{*} = 60$.  From top to bottom circle points correspond to $ \beta = 0.0085 $, $ \beta = 0.0368 $, $ \beta = 0.2572 $, $ \beta = 0.5013 $, $ \beta = 0.3894 $ and $ \beta = 0.6420 $ respectively. From top to bottom star points correspond to $ \beta = 0.0053 $, $ \beta = 0.0230 $, $ \beta = 0.1714 $, $ \beta = 0.3342 $, $ \beta = 0.2596 $ and $ \beta = 0.5707 $ respectively.} 
\end{figure}

%%%%%%%%%%%%%%%%%%%%%%%%%%%%%

\section{Conclusion}

In this paper, we have considered GCG as a natural candidate for the inflation. In this model, the equation of state for GCG starts with  $w=-1$ behaviour and leads to inflation. With time, the equation of state naturally evolves towards $w=0$ dust behaviour and inflation ends. Subsequently we study a canonical scalar field theory that represents the GCG behaviour.

While studying the primordial fluctuations in this model, we show that in Einstein gravity, GCG is not suitable for generating the required PPS as one needs a fairly large value for $N_{*}$ ($N_{*} = 217$), the required e-folding at horizon exit, which is incompatible with theoretical constraint $50<N_{*}<60$.  Next we consider the GCG inflationary model in RS type five dimensional brane world scenario where the Einstein equation gets a correction term due to the presence of higher dimension. In this set up, we show that GCG works perfectly as a slow-roll inflationary model. We obtain the constraints on the model parameters using the bounds on $n_{s}, A_{s}$ and $r$ as obtained by Planck and BICEP2. The inflationary energy scale in our model is around $5 \times 10^{15}$ GeV, one order less than the GUT scale. The value of the five-dimensional Planck mass is around $10^{-2} M_{Pl}$ in our model.

As any other slow-roll inflationary model with power-law type scalar PPS, GCG model is also in tension with combined Planck+BICEP2 data primarily due to the fact that large contribution from gravitational wave as measured by BICEP2, can not explain the suppression of power at large scales as observed by Planck. We show that by allowing a general dark energy equation of state given by CPL parametrization, one can ease this tension as a general dark energy behaviour may allow suppression of power at low $l$ even with $r=0.2$. By calculating the best fit likelihood values for $\Lambda$CDM and CPLDE model, we show that CPLDE model with a power law type scalar PPS is a better fit to the joint Planck+WP+BICEP2 data compared to a $\Lambda$CDM with similar PPS. Hence allowing a deviation from concordance $\Lambda$CDM model may save the simple slow-roll inflationary models with power-law type scalar PPS. Finally we do a full MCMC analysis using COSMOMC with a CPLDE model using Planck+WP+BICEP2 data and obtain the revised estimate for $n_{s}$ and $r$ as well as our model parameters $\alpha$ and $\beta$. 

It will be interesting to extend our work to phantom scalar inflationary models as previously studied in \cite{phinf} and we hope to address this issue in future.

\section*{Acknowledgements}
The authors would like to thank Dhiraj Kumar Hazra for extensive discussions and valuable comments. We also acknowledge the use of publicly available codes, CAMB and COSMOMC. A.A.S. acknowledges the funding from SERC, Dept. of Science and Technology, Govt. of India through the research project SR/S2/HEP-43/2009.  B.R.D. thanks CSIR, Govt. of India for financial support through JRF scheme (No:09/466(0157)/2012-EMR-I).  S.K. thanks the UGC, Govt. of India for financial support through SRF scheme.

\end{document}